# A Semi-Analytical Approach for State-Space Electromagnetic Transient Simulation Using the Differential Transformation

Min Xiong, *Student Member, IEEE*, Kaiyang Huang, *Student Member, IEEE*, Yang Liu, *Member, IEEE*, Rui Yao, *Senior Member, IEEE*, Kai Sun, *Senior Member, IEEE,* Feng Qiu, *Senior Member, IEEE*

*Abstract*—**Electromagnetic transient (EMT) simulation is a crucial tool for power system dynamic analysis because of its detailed component modeling and high simulation accuracy. However, it suffers from computational burdens for large power grids since a tiny time step is typically required for accuracy. This paper proposes an efficient and accurate semi-analytical approach for state-space EMT simulations of power grids. It employs high-order semi-analytical solutions derived using the differential transformation from the state-space EMT grid model. The approach incorporates a proposed variable time step strategy based on equation imbalance, leveraging structural information of the grid model, to enlarge the time step and accelerate simulations, while high resolution is maintained by reconstructing detailed fast EMT dynamics through an efficient dense output mechanism. It also addresses limit-induced switches during large time steps by using a binary search-enhanced quadratic interpolation algorithm. Case studies are conducted on EMT models of the IEEE 39-bus system and a synthetic 390-bus system to demonstrate the merits of the new simulation approach against traditional methods.**

## I. INTRODUCTION

AS powerful tools for power system dynamic and stability analysis, numerical power system simulations include electromechanical transient simulations and electromagnetic transient (EMT) simulations. The former adopt a phasor model for the simulated power system, assuming a synchronous system frequency and computing only positive-sequence values of voltages and currents. The simulation time steps are around milliseconds to focus on electro-mechanical dynamics of generators and motors. The latter, EMT simulations, compute three-phase instantaneous voltages and currents varying at actual frequencies, taking into account the fast dynamics caused by the interactions between the electrical fields of capacitors and magnetic fields of inductors. Thus, EMT simulations enable a more detailed analysis of fast phenomena such as harmonics, overvoltage, short-circuit current, and other fast dynamics. However, EMT simulations suffer from high computational costs due to the tiny time steps in μs [1]-[3].

To accelerate EMT simulations, mainstream solutions adopt hybrid simulation or parallel computing [4]-[13]. Reference [5] proposed a hybrid simulation protocol to reduce the computational burden, in which only parts of the system requiring detailed dynamic information are simulated using an EMT model, and the rest is studied by phasor domain simulation. This approach has been further investigated by [6]-[9] to validate its feasibility. Parallel computing has also been adopted to boost EMT simulation [10]-[13]. Ref. [14] proposes a state-space nodal method, which discrete state-space equations of subsystems to nodal equations and enables flexible interfacing of subsystems in nodal and state-space representations.

Some other approaches for accelerating EMT simulations focus on enlarging the time step to reduce the number of simulation steps. Related approaches include DQ0 transformation, frequency shift, and dynamic phasor methods. Because balanced three-phase instantaneous variables are transformed to DC components under DQ0 transformation, the dynamics can be simulated more efficiently. However, both the DQ0 transformation and frequency shift methods are inefficient for unbalanced conditionssss or the presence of high-order harmonics. The dynamic phasor method approximates the original waveform using $K$ time-varying Fourier series, but it requires careful selection of $K$ to accurately reflect the correct system behaviors. The number of variables and equations will rise with the increase of $K$ and thus decrease the efficiency [15]-[18].

In recent years, semi-analytical solutions (SAS's) of power systems have been applied to speed up electromechanical transient simulations. Each SAS is an approximate analytical solution for a nonlinear system modeled by ordinary differential equations or differential-algebraic equations and can be derived by mathematical methods such as the Adomian Decomposition, Holomorphic Embedding, Differential Transformation (DT), and Homotopy Analysis [19]-[25]. Among these methods, the DT provides a highly accurate approximation of a true solution by using an ultra-high-order SAS in the form of a power series of time, and allows deriving the SAS terms from low orders to arbitrary high orders in an efficient, recursive manner. Besides power system simulations, the DT has proven successful in other fields [26]-[28].

For the first time, this paper proposes a DT-based semi-analytical EMT simulation approach while existing semi-analytical

This work was supported in part by the ERC Program of the NSF and U.S. DOE under grant EEC-1041877 and in part by the Advanced Grid Modeling program of U.S. DOE Office of Electricity under grant DE-OE0000875.

M. Xiong, K. Huang, and K. Sun are with the University of Tennessee, Knoxville, TN, 37996 USA. (e-mail: mxiong3@vols.utk.edu, khuang12@vols.utk.edu, kaisun@utk.edu).

Y. Liu and F. Qiu are with the Argonne National Laboratory, Lemont, IL, 60439 USA. (e-mail: yang.liu100@anl.gov, fqiu@anl.gov).

R. Yao was with the Argonne National Laboratory, Lemont, IL, 60439 USA and is now with X Development LLC, Mountain View, CA 94043. (e-mail: yaorui.thu@gmail.com).

simulation methods for power systems have focused on phasor models for electromechanical transient simulations. Based on EMT state-space equations, the proposed approach derives SASs for network components represented by $R$-$L$-$C$ circuits, for the Voltage-Behind-Reactance (VBR) synchronous generator model, and for typical generator controllers. Then, a multi-stage simulation strategy is proposed, enabling variable, larger time steps for acceptable truncation errors estimated by equation imbalance. Whenever fast dynamics become dominant, this approach can automatically adjust the step size for convergent computations and then adaptively increase the step size thereafter for acceleration. As an analytical function of time, the SAS allows its values to be calculated at any time instants directly through its evaluation to provide high-resolution results.

Also, the proposed approach can accurately handle system switches during a simulation caused by limit violations. For EMT simulations, finding the limit violation moment is important for precise and correct simulation results, but the traditional linear interpolation algorithm for small-step EMT simulations may become inaccurate at time steps enlarged by SAS. This paper proposes a binary search-enhanced quadratic-interpolation algorithm with the SAS to precisely locate the moment of a switch when any limit is hit, which has not been fully addressed by previous SAS-based simulation methods. The high-order SAS derived from the DT allows adjusting the step size immediately when a switch occurs to correct the simulation.

The rest of this paper is organized as follows. Section II introduces the proposed DT-based semi-analytical EMT simulation approach and the set of transformation rules for linear and nonlinear state-space EMT models. Section III derives the SAS for each component model. Section IV presents and illustrates a multistage simulation strategy incorporating a proposed variable time step algorithm, along with the dense output mechanism. Section V proposes a binary search-enhanced quadratic interpolation algorithm for limit violation detection. Then, tests on EMT models of the IEEE 39-bus system and a synthetic 390-bus system are conducted respectively in Section VI and Section VII. Finally, conclusions are drawn in section VIII.

## II. SAS AND DIFFERENTIAL TRANSFORMATION RULES FOR EMT SIMULATION

An EMT simulation needs to solve an initial value problem for the power system model. The $N$-th order SAS for the initial value problem (1) takes the form (2) [20]-[23]:

$$\frac{dx(t)}{dt} = f(x(t), t)$$
$$x\big|_{t=0} = x[0] \tag{1}$$

$$x(t) = \sum_{k=0}^{\infty} x[k]t^k \approx \sum_{k=0}^{N} x[k]t^k, \qquad t < T_x \tag{2}$$

where $T_x$ is the convergence region; $x[0]$ is the initial value of $x$; $x[k]$ indicates coefficients of the SAS series; $k$ is the order of series terms; and $N$ is the highest order of SAS terms.

Compared with low-order numerical methods such as the 4th-order Runge-Kutta method, the SAS offers higher-order approximations to achieve a bigger convergence region and allow a larger time step while maintaining the same level of simulation accuracy.

The $k^{th}$-order DT of a smooth function $g(t)$ at $t=t_0$ is defined in [26] as:

$$g[k] = \frac{1}{k!}\left[\frac{d^k g(t)}{dt^k}\right]_{t=t0} \tag{3}$$

Thus, from (1)-(3), a differential equation (4) involving smooth functions $f(t)$, $g(t)$, and $h(t)$ is transformed into (5) about $f[\mathrm{k}]$, $g[k]$ and $h[k]$ ($k$=0, 1, ..., $N$), which are their DTs of the $k$-th order [26]-[27]:

$$\frac{dg(t)}{dt} = f(t)$$
$$f(t) = g(t)h(t) \tag{4}$$

$$\sum_{k=0}^{N}(k+1)g[k+1](t-t_0)^k = \sum_{k=0}^{N} f[k](t-t_0)^k$$
$$\sum_{k=0}^{N} f[k](t-t_0)^k = \sum_{k=0}^{N}\sum_{m=0}^{k}(g[m]h[k-m])(t-t_0)^k$$
$$\Rightarrow \qquad (k+1)g[k+1] = f[k] \tag{5}$$
$$f[k] = \sum_{m=0}^{k} g[m]h[k-m]$$

By calculating DTs in such a recursive manner, the SAS of any order for (1) can be derived in the form of (2).

Some transformation rules for common linear and nonlinear functions are given in Table I [26]-[28], which facilitate efficient, recursive derivation of a high-order SAS.

TABLE I
DIFFERENTIAL TRANSFORM RULES

| Original function | Differential Transformation |
|---|---|
| #1 $f(t) = c$<br>$c$ is constant | $f[k] = c\eta[k], \ \eta[k] = \begin{cases} 1 & k=0 \\ 0 & k \neq 0 \end{cases}$ |
| #2 $f(t) = cg(t)$ | $f[k] = cg[k]$ |
| #3 $f(t) = g(t) \pm h(t)$ | $f[k] = g[k] \pm h[k]$ |
| #4 $f(t) = g(t)h(t)$ | $f[k] = \sum_{m=0}^{k} g[m]h[k-m]$ |
| #5 $f(t) = \dfrac{dg(t)}{dt}$ | $f[k] = (k+1)g[k+1]$ |
| #6 $f(t) = \sin(h(t))$<br>$g(t) = \cos(h(t))$ | $f[k] = \sum_{m=0}^{k-1}\dfrac{k-m}{k} g[m]h[k-m]$<br>$g[k] = -\sum_{m=0}^{k-1}\dfrac{k-m}{k} f[m]h[k-m]$ |
| #7 $s(t) = g^2(t) + h^2(t)$<br>$f(t) = \sqrt{s(t)}$ | $s[k] = \sum_{m=0}^{k}(g[m]g[k-m] + h[m]h[k-m])$<br>$f[k] = \dfrac{1}{2f_0}(s[k] - \sum_{m=1}^{k-1} f[m]f[k-m])$ |

$$\begin{bmatrix} z_0(t) \\ z_d(t) \\ z_q(t) \end{bmatrix} = \underbrace{\frac{2}{3}\begin{bmatrix} \dfrac{1}{2} & \dfrac{1}{2} & \dfrac{1}{2} \\ \cos\theta & \cos(\theta - \dfrac{2\pi}{3}) & \cos(\theta + \dfrac{2\pi}{3}) \\ -\sin\theta & -\sin(\theta - \dfrac{2\pi}{3}) & -\sin(\theta + \dfrac{2\pi}{3}) \end{bmatrix}}_{P_{park}} \begin{bmatrix} z_a(t) \\ z_b(t) \\ z_c(t) \end{bmatrix} \tag{6}$$

In EMT simulation, the Park transformation (6) and its inverse transformation are required to interface variables of a generator in the DQ0 frame with the three-phase network in the *abc* frame. To deal with matrix multiplications, the following *Proposition* 1 is introduced, based on rules in Table I.

*Proposition 1*: For any three smooth matrix functions $F(t)$, $G(t)$ and $H(t)$ given in (7) that satisfy $F(t)=G(t)H(t)$ for $t \geq 0$, the rule in (8) holds for DTs of their matrix elements.

$$F(t) = [f_{ij}(t)] \in \mathbb{R}^{N_0 \times N_2}$$
$$G(t) = [g_{ij}(t)] \in \mathbb{R}^{N_0 \times N_1} \quad (7)$$
$$H(t) = [h_{ij}(t)] \in \mathbb{R}^{N_1 \times N_2}$$

$$f_{ij}[k] = \sum_{n=1}^{N_1} \sum_{m=0}^{k} g_{in}[m] h_{nj}[k-m] \quad (8)$$

where $f_{ij}[k]$, $g_{ij}[k]$, and $h_{ij}[k]$ are the $i$-th row $j$-th column elements, $N_0$, $N_1$ and $N_2$ are positive integers on matrix dimensions, and $[k]$, $[m]$, and $[k-m]$ indicate the orders of DTs.

## III. DT OF STATE-SPACE REPRESENTED EMT MODELS

This section presents state-space EMT models of power system components along with their SASs derived by the DT.

### A. Synchronous Generator

The volage-behind-reactance (VBR) model from [29] is applied in this paper for EMT simulation purpose. Eq. (9)-(11) give the VBR model of a round rotor generator, comprising the swing equation (9), the rotor flux equation (10), and the stator phase current equation that interface with the network (11) [30]:

$$\frac{d\delta}{dt} = \Delta\omega_r$$
$$\frac{d\Delta\omega_r}{dt} = \frac{\omega_0}{2H}(p_m - p_e - D\frac{\Delta\omega_r}{\omega_0}) \quad (9)$$

$$\frac{d\lambda_{fd}}{dt} = e_{fd} - \frac{r_{fd}}{L_{lf}}(\lambda_{fd} - \lambda_{ad})$$
$$\frac{d\lambda_{1d}}{dt} = -\frac{r_{1d}}{L_{1dl}}(\lambda_{1d} - \lambda_{ad})$$
$$\frac{d\lambda_{1q}}{dt} = -\frac{r_{1q}}{L_{1ql}}(\lambda_{1q} - \lambda_{aq}) \quad (10)$$
$$\frac{d\lambda_{2q}}{dt} = -\frac{r_{2q}}{L_{2ql}}(\lambda_{2q} - \lambda_{aq})$$

$$\frac{di_{abc}}{dt} = -\frac{1}{L''_{abc}}(v_{abc} + R_s i_{abc} - P^{-1}_{park} v''_{0dq} + \frac{dL''_{abc}}{dt} i_{abc}) \quad (11)$$

In (9), $\delta$, $\Delta\omega_r$, $H$, $D$, $p_m$, and $p_e$ are the rotor angle, rotor speed deviation, inertial constant, damping coefficient, mechanical power, and electrical power, respectively, and $\omega_0$ is the nominal frequency of the system. In (10), $\lambda_{fd}$, $\lambda_{1d}$, $\lambda_{1q}$, and $\lambda_{2q}$ are flux linkages of the filed winding, $d$-axis damper winding, $q$-axis first damper winding, $q$-axis second damper winding, respectively; $r_{fd}$, $r_{1d}$, $r_{1q}$, and $r_{2q}$ are resistances of the four windings; $L_{fdl}$, $L_{1dl}$, $L_{1ql}$, and $L_{2ql}$ are leakage inductance of the four windings; $e_{fd}$ is field voltage. In (11), $i_{abc}$ and $v_{abc}$ are three-phase generator terminal current and voltage that interface with the grid through current injection [30]; $R_s$ is the constant stator resistance. The subtransient inductance matrix $L''_{abc}$ in (11) is:

$$L''_{abc} = \begin{bmatrix} L_S(2\theta) & L_M(2\theta - \frac{2\pi}{3}) & L_M(2\theta + \frac{2\pi}{3}) \\ L_M(2\theta - \frac{2\pi}{3}) & L_S(2\theta + \frac{2\pi}{3}) & L_M(2\theta) \\ L_M(2\theta + \frac{2\pi}{3}) & L_M(2\theta) & L_S(2\theta - \frac{2\pi}{3}) \end{bmatrix} \quad (12)$$

where,

$$L_S(\cdot) = L_{ls} + \frac{1}{3}(L_0 - L_d + L''_{ad} + L''_{aq}) + \frac{1}{3}(L''_{ad} - L''_{aq})\cos(\cdot)$$
$$L_M(\cdot) = \frac{1}{6}(2L_0 - 2L_d - L''_{ad} - L''_{aq}) + \frac{1}{3}(L''_{ad} - L''_{aq})\cos(\cdot) \quad (13)$$
$$L''_{ad} = L_{ad} // L_{fdl} // L_{1dl}, \quad L''_{aq} = L_{aq} // L_{1ql} // L_{2ql}$$

Here, $L_{ls}$ and $L_0$ are leakage inductance and zero sequence inductance, respectively, and "$//$" means the parallel connection of inductances.

The $d$-$q$ axes subtransient voltages in (11) are given by:

$$v''_d = -(\frac{r_{fd}}{L^2_{fdl}} + \frac{r_{1d}}{L^2_{1dl}})L''^2_{ad}i_d - [\frac{r_{fd}L''_{ad}}{L^2_{fdl}}(1 - \frac{L''_{ad}}{L_{fdl}}) - \frac{L''^2_{ad}}{L^2_{1dl}}\frac{r_{1d}}{L_{1dl}}]\lambda_{fd}$$
$$+ [\frac{L''^2_{ad}}{L^2_{fdl}}\frac{r_{fd}}{L_{1dl}} - \frac{r_{1d}L''_{ad}}{L^2_{1dl}}(1 - \frac{L''_{ad}}{L_{1dl}})]\lambda_{1d} - L''_q(\frac{\lambda_{1q}}{L_{1ql}} + \frac{\lambda_{2q}}{L_{2ql}})\omega_r + \frac{L''_{ad}}{L_{fdl}}e_{fd} \quad (14)$$

$$v''_q = -(\frac{r_{1q}}{L^2_{1ql}} + \frac{r_{2q}}{L^2_{2ql}})L''^2_{aq}i_q - [\frac{r_{1q}L''_{aq}}{L^2_{1ql}}(1 - \frac{L''_{aq}}{L_{1ql}}) - \frac{L''^2_{aq}}{L^2_{2ql}}\frac{r_{2q}}{L_{2ql}}]\lambda_{1q}$$
$$+ [\frac{L''^2_{aq}}{L^2_{1ql}}\frac{r_{1q}}{L_{2ql}} - \frac{r_{2q}L''_{aq}}{L^2_{2ql}}(1 - \frac{L''_{aq}}{L_{2ql}})]\lambda_{2q} + L''_d(\frac{\lambda_{fd}}{L_{fdl}} + \frac{\lambda_{1d}}{L_{1dl}})\omega_r \quad (15)$$

The $d$-$q$ axes flux linkages $\lambda_{ad}$ and $\lambda_{aq}$ are:

$$\lambda_{ad} = (-i_d + \frac{\lambda_{fd}}{L_{fdl}} + \frac{\lambda_{1d}}{L_{1dl}})L''_{ad}$$
$$\lambda_{aq} = (-i_q + \frac{\lambda_{1q}}{L_{1ql}} + \frac{\lambda_{2q}}{L_{2ql}})L''_{aq} \quad (16)$$

In the Park transformation (6), $\theta$ satisfies:

$$\frac{d\theta}{dt} = \omega_r \quad (17)$$

The electrical power depends on the number of poles $n_p$:

$$p_e = \frac{3n_p}{4}\omega_r(\lambda_{ad}i_q - \lambda_{aq}i_d) \quad (18)$$

Applying the rules in Table I to (9)-(18) yields the corresponding DT equations as illustrated in (19)-(23) for variables $\delta$, $\Delta\omega_r$, $\lambda_{fd}$, $i_{abc}$, $v''_q$, and $p_e$.

$$(k+1)\delta[k+1] = \Delta\omega_r[k]$$
$$(k+1)\Delta\omega_r[k+1] = \frac{\omega_0}{2H}(p_m[k] - p_e[k] - D\frac{\Delta\omega_r[k]}{\omega_0}) \quad (19)$$

$$(k+1)\lambda_{fd}[k+1] = e_{fd}[k] - \frac{r_{fd}}{L_{lf}}(\lambda_{fd}[k] - \lambda_{ad}[k]) \quad (20)$$

$$(k+1)i_{abc}[k+1] = -L''^{-1}_{abc}(v_{abc}[k] + R_s i_{abc}[k] - v''_{abc}[k]) \quad (21)$$

where, $v''_{abc}[k]$ in (21) can be derived from applying the *Proposition 1* to $v''_{abc} = P^{-1}_{park}v''_{0dq}$.

$$v''_q[k+1] = -(\frac{r_{1q}}{L^2_{1ql}} + \frac{r_{2q}}{L^2_{2ql}})L''^2_{aq}i_q[k+1] - [\frac{r_{1q}L''_{aq}}{L^2_{1ql}}(1 - \frac{L''_{aq}}{L_{1ql}}) - \frac{L''^2_{aq}}{L^2_{2ql}}\frac{r_{2q}}{L_{2ql}}]\lambda_{1q}[k+1]$$
$$+ [\frac{L''^2_{aq}}{L^2_{1ql}}\frac{r_{1q}}{L_{2ql}} - \frac{r_{2q}L''_{aq}}{L^2_{2ql}}(1 - \frac{L''_{aq}}{L_{2ql}})]\lambda_{2q}[k+1] + L''_d\sum_{m=0}^{k+1}(\frac{\lambda_{fd}[m]}{L_{fdl}} + \frac{\lambda_{1d}[m]}{L_{1dl}})\omega_r[k+1-m] \quad (22)$$

$$p_e[k+1] = \frac{3n_p}{4}\sum_{m=0}^{k+1}\sum_{n=0}^{m}\omega_r[k+1-m](\lambda_{ad}[m-n]i_q[n] - \lambda_{aq}[m-n]i_d[n]) \quad (23)$$

where, $i_d[n]$ and $i_q[n]$ in (23) are obtained by applying the *Proposition 1* to $i_{0dq} = P_{park} i_{abc}$.

### B. Controllers of a Generator

Also consider the models of the exciter, turbine, and governor of a synchronous generator. Without loss of generality, the TGOV1 turbine-governor model in [31] and SEXS exciter model in [32] are adopted in this paper as described below.

The TGOV1 model is shown in (24):

$$
\begin{aligned}
&P_{min} \leq p_1 \leq P_{max} \\
&\frac{dp_1}{dt} = \frac{1}{T_1}(\frac{1}{R_G}(p_{ref} - \Delta\omega_r) - p_1) \\
&\frac{dp_2}{dt} = \frac{1}{T_3}(T_2\frac{dp_1}{dt} + p_1 - p_2) \\
&p_m = p_2 - D_t\Delta\omega_r
\end{aligned}
\tag{24}
$$

where $p_1$ and $p_2$ are intermediate state variables; $p_{ref}$ is the reference power, and others are control parameters. The corresponding DT is given as (25). In addition, due to the limits, if initial value $p_1[0] \geq P_{max}$ and $dp_1/dt \geq 0$, or if $p_1[0] \leq P_{min}$ and $dp_1/dt \leq 0$, then $p_1[k+1]=0$.

$$
\begin{aligned}
&P_{min} \leq p_1[0] \leq P_{max} \\
&(k+1)p_1[k+1] = \frac{1}{T_1}(\frac{1}{R_G}(\eta[k]p_{ref} - \Delta\omega_r[k]) - p_1[k]) \\
&(k+1)p_2[k+1] = \frac{1}{T_3}((k+1)T_2 p_1[k+1] + p_1[k] - p_2[k]) \\
&p_m[k+1] = p_2[k+1] - D_t\Delta\omega_r[k+1]
\end{aligned}
\tag{25}
$$

The SEXS exciter is described as:

$$
\begin{aligned}
&E_{min} \leq E_{fd} \leq E_{max} \\
&\frac{dE_{fd}}{dt} = \frac{1}{T_E}(k_E v_1 - E_{fd}) \\
&\frac{dv_1}{dt} = \frac{1}{T_B}(v_{ref} - T_A\frac{dv_1}{dt} - v_t - v_1)
\end{aligned}
\tag{26}
$$

where $v_{ref}$ is voltage regulator reference; $E_{fd}=e_{fd}L_{ad}/R_{fd}$; $v_1$ is an intermediate variable; $v_t$ is terminal voltage; $K_E$, $T_E$, $T_A$, $T_B$ are exciter control parameters. Its DT is given as (27). In addition, if $E_{fd}[0] \geq E_{max}$ and $dE_{fd}/dt \geq 0$, or if $E_{fd}[0] \leq P_{min}$ and $dE_{fd}/dt \leq 0$, then $E_{fd}[k+1]=0$.

$$
\begin{aligned}
&E_{min} \leq E_{fd}[0] \leq E_{max} \\
&(k+1)E_{fd}[k+1] = \frac{1}{T_E}(k_E v_1[k] - E_{fd}[k]) \\
&(k+1)v_1[k+1] = \frac{1}{T_B}(\eta[k]v_{ref} - (k+1)T_A v_1[k+1] - v_t[k] - v_1[k])
\end{aligned}
\tag{27}
$$

### C. Network Components

Transmission lines are modeled as Π sections in this work [33]. Other network components such as transformers, constant impedance loads, and fixed shunts, are modeled by $R$-$L$-$C$ circuits. Two typical circuits, including the series resistor-inductor circuits and grounding capacitor circuits, are shown in Fig. 1 as simple examples:

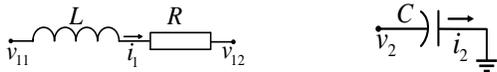

(a) series resistor-inductor circuit      (b) grounding capacitor circuit

Fig. 1. $R$-$L$-$C$ circuit examples

The differential equations and corresponding DTs of the resistor-inductor circuit and grounding capacitor circuit in Fig. 1 are respectively given by (28)-(29) and (30)-(31):

$$
\frac{di_1}{dt} = \frac{v_{11} - v_{12} - Ri_1}{L}
\tag{28}
$$

$$
(k+1)i_1[k+1] = \frac{v_{11}[k] - v_{12}[k] - Ri_1[k]}{L}
\tag{29}
$$

$$
\frac{dv_2}{dt} = \frac{i_2}{C}
\tag{30}
$$

$$
(k+1)v_2[k+1] = \frac{i_2[k]}{C}
\tag{31}
$$

Note that other $R$-$L$-$C$ branch types can also be easily considered and are not presented in detail here.

*Remark:* Using the DT rules, the SAS of other types of exciters, governors, and stabilizers can be derived. Also, nonlinear models such as the piecewise-linear magnetic saturation of synchronous generator [34] can be transformed by using the DT.

### D. Procedure for Calculating SAS of the System

Based on the DT equations derived in the last section, first separate all state variables into three groups:

$x_1(t)=[\delta, \Delta\omega_r, \lambda_{fd}, \lambda_{1d}, \lambda_{1q}, \lambda_{2q}, \theta, i_{abc}, p_1, p_2, E_{fd}, v_1]^T$

$x_2(t)=[i_{net}, v_{net}]^T$

$x_3(t)=[v_{0dq}, i_{0dq}, v''_{0dq}, \lambda_{ad}, \lambda_{aq}, p_e, p_m]^T$

where $x_1$ and $x_3$ includes all state variables of synchronous generators and related controls described by differential equations and non-differential equations, respectively; $x_2$ includes all network state variables described by differential equations; $i_{net}$ includes all network inductor currents; $v_{net}$ includes all network capacitance voltages.

Then, the simulation process for each time step is conducted following the **Algorithm 1** summarized below.

---

**Algorithm 1**

**Input**: initial state $x_1[0]$, $x_2[0]$, $x_3[0]$ at $t=t_0$, time step $\Delta t$, SAS order $N$

**Output**: SAS coefficients $x_1[1:N]$, $x_2[1:N]$, $x_3[1:N]$, and states $x_1(t)$, $x_2(t)$, $x_3(t)$ at $t= t_0+\Delta t$

1:   $k=0$
2:   **While** $k<N$
3:      $k=k+1$
4:      Calculate $x_1[k]$ using (19), (20), (21), (25), (27), etc.
5:      Calculate $x_2[k]$ using (29), (31).
6:      Calculate $x_3[k]$ using (22), (23), (25), etc.
7:   **End while**
8:   Calculate values of $x_1$, $x_2$, and $x_3$ by using:
     $x_i(t+\Delta t) = \sum_{k=0}^{N} x_i[k](\Delta t)^k \quad i=1,2,3$

---

## IV. SIMULATION STRATEGIES

### A. Multistage Simulation Strategy

Because the SAS is not an exact solution, its error can be tolerated only within a limited time step. The initial value problem to be solved for simulation can evaluate an SAS by a multistage strategy as depicted in Fig. 2 over consecutive time steps that make the desired simulation period. At the end of each time step, the end values of state variables are used as the initial values for the next time step to evaluate the SAS, which can be derived ahead of time or together with the simulation.

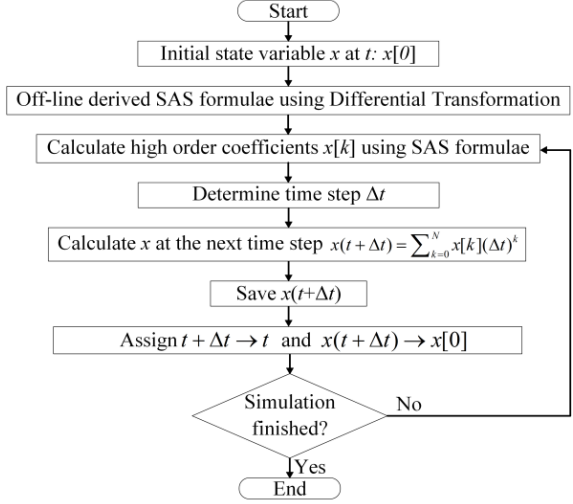

Fig. 2. Flow chart of the multistage simulation strategy using DT

### B. Variable Time Step Simulation Strategy

The simulation can be conducted at a fixed or variable time step. To improve the convergence and efficiency of simulation using a SAS over larger time steps, a variable-time-step strategy is proposed and incorporated as follows.

Suppose a linear state-space EMT model for the system [35][36]:

$$\dot{x} = Ax + Bu \qquad (32)$$

where $x$ is network state vector; $u$ is input vector; $A$ is state matrix, and $B$ is input matrix. Substitute the SAS for $x$ and calculate the absolute imbalance in per unit due to its:

$$E(\Delta t) = \|\dot{x}_{SAS} - Ax_{SAS} - Bu\|_\infty \qquad (33)$$

where,

$$x_{SAS} = \sum_{k=0}^{N} x[k](\Delta t)^k \qquad (34)$$
$$(k+1)x[k+1] = Ax[k] + Bu[k]$$

Substituting (34) into (33) yields:

$$E(\Delta t) = \|Ax[N] + Bu[N]\|_\infty (\Delta t)^N \qquad (35)$$

The imbalance $E(\Delta t)$ can serve as an indicator for determining the termination of the present time interval and the initiation of the subsequent one. Thus, a variable time step is enabled for an appropriate balance between the step size and accuracy [22][23]. In practice, the variable length of each time step is determined by a pre-determined imbalance threshold $\varepsilon_E$, typically around $1\times10^{-2}$ per unit for EMT simulation.

$$E(\Delta t) \le \varepsilon_E \ \Rightarrow \ \Delta t \le \Delta t_{max} \qquad (36)$$

### C. Dense Output Mechanism

Because the SAS utilizes high-order approximations to enlarge the time steps, some detailed fast dynamics such as overvoltage and over current may be missed. Thus, obtaining dense output is of significance in such cases.

As shown in (2) and (34), with derived SAS coefficients, the SAS expression is a function of time. Thus, the value of state variables at the next time step can be calculated by substituting the time step $\Delta t$. Furthermore, it is flexible to calculate values at multiple instants $t_i$ within the time step $\Delta t$, as illustrated in (37) and Fig. 3 below.

$$x(t_n) \approx \sum_{k=0}^{N} x[k]t_n^k \qquad (37)$$
$$t_n < t_0 + \Delta t \quad n=1,2,3,\cdots$$

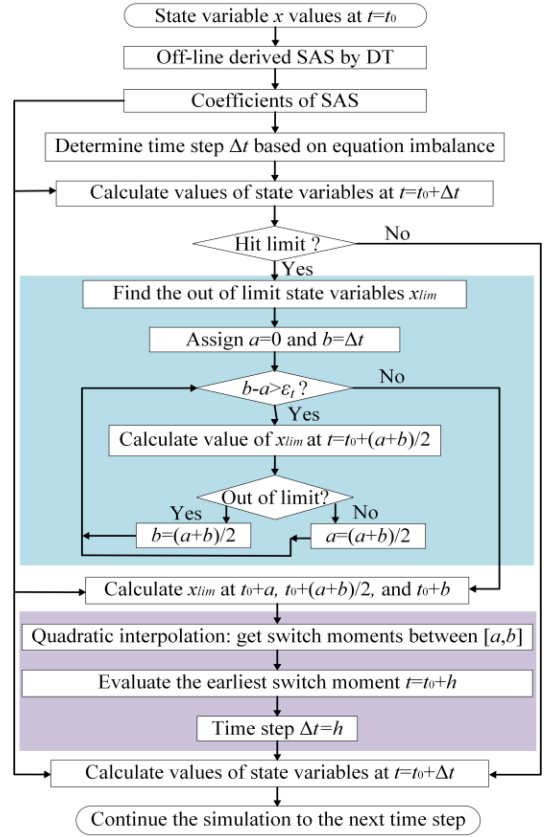

Fig. 3. Illustration of dense output under large time step simulations

## V. LIMIT VIOLATION DETECTION ALGORITHM

In EMT simulations, the time instants when limits are reached are typical of interest and importance to ensure the correctness of results. The timings of such switches are unknown ahead of the simulation. In commercial software like PSCAD, if a switch occurs during a small time step, its timing is estimated via linear interpolation by using values at the previous and present steps, followed by a re-computation for the present step, starting from the switching moment [37][38]. However, when a significantly increased time step is used by an SAS, linear interpolation may yield inaccurate results.

### A. Proposed Limit Violation Detection Algorithm

Fig. 4. Flow chart of the switch detection algorithm for limit violations

To address the switch caused by limit violations, a switch detection algorithm utilizing SAS coefficients is proposed as depicted in Fig. 4. Thanks to high-order SAS coefficients, state variable values at any moment within a time step can be calculated efficiently following (2) with negligible arithmetic operations. Moreover, to avoid an exhaustive and time-consuming process of checking every tiny time step at 1 μs or less, the earliest small time interval $\varepsilon_t$, e.g., 10-20 μs or smaller, in which any limit violation happened is identified by a binary search approach as illustrated by steps in the blue area of Fig. 4 on hitting upper limits. To deal with lower limits, the algorithm requires only a swap in the updates of $a$ and $b$. Then, as indicated by steps in the purple area, as $a$ and $b$ are close enough, quadratic interpolation is used to accurately detect the earliest switch moment where the current time step should be finished.

## VI. CASE STUDY ON THE IEEE 39-BUS SYSTEM

To evaluate the efficacy and validity of the proposed approaches, case studies on the IEEE 39-bus system [39] are conducted and presented in this section.

### A. Benchmark Against PSCAD

The SAS-based simulation approach is implemented in MATLAB and benchmarked with the commercial tool PSCAD. Consider a contingency in which the system starts from its steady state at $t=0$ s, and then has a three-phase grounding fault on Bus 10 at $t=1$ s lasting for 0.2 second without losing any network component. The results from the proposed approach using MATLAB and from PSCAD are compared in Fig. 5-9, which match well. Some tiny mismatch between the two results is observed, which is acceptable and may be caused by unknown implementation details with PSCAD.

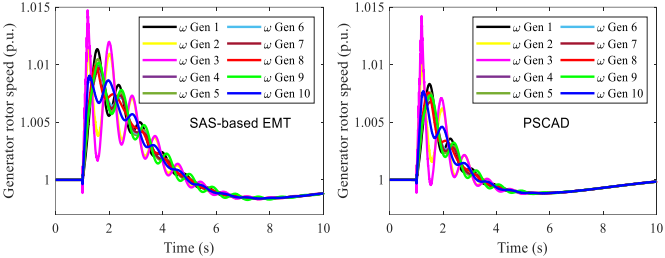

Fig. 5. Simulation results of generator rotor speed

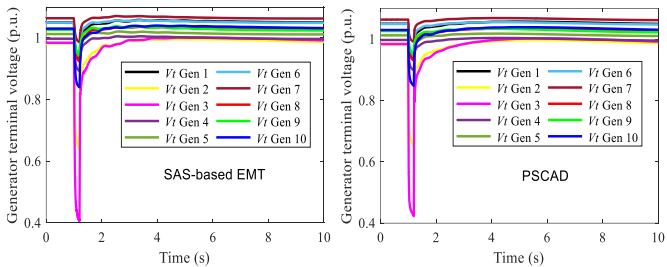

Fig. 6. Simulation results of generator terminal voltage

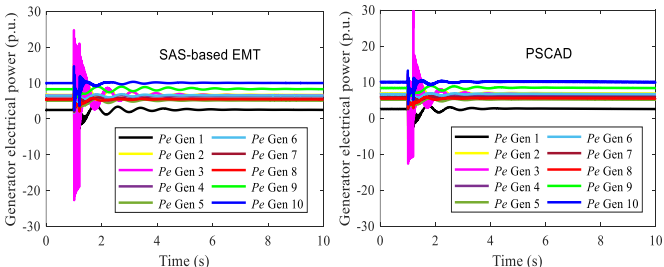

Fig. 7. Simulation results of generator electrical power

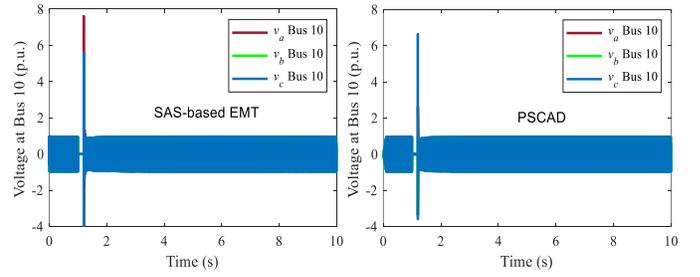

Fig. 8. Simulation results of three-phase voltage at Bus 10

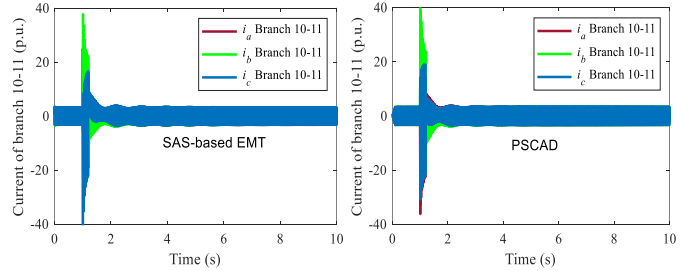

Fig. 9. Simulation results of three-phase current on branch 10-11

### B. Evaluating the SAS-Based Approach with Different Orders

The performance of the proposed simulation approach using the SASs of different orders with a variable time step is tested and compared with traditional numerical solvers on the state-space model.

Without loss of generality, three different cases including load tripping, bus grounding, and generator tripping are simulated: Case 1 disconnects the load at bus 4 at $t=1$ s; Case 2 adds a three-phase grounding fault to bus 10 at $t=1$ s lasting for 0.2 second; Case 3 disconnects the generator at bus 36.

A higher-order SAS allows longer and fewer time steps, but its elevated complexity increases the computation burden for each time step. Thus, there exists an optimal SAS order with the best performance. Using different orders, the average time steps and time costs of SAS for a 1-s simulation are summarized in Fig. 10 for three cases. The benchmark results are from the $4^{th}$-order Runge-Kutta method using a 1 μs time step. The simulation tests are carried out on a desktop computer equipped with an Intel Core i7-6700 3.4 GHz CPU and 16GB RAM, using MATLAB. To ensure accuracy and convergence, the errors of bus voltages are limited to less than 0.01 p.u.

To notice, MATLAB is an interpreted rather than a compiled environment, having hidden overheads. To accelerate the simulation speed, the MATLAB codes are converted into MEX files and compiled to enable higher efficiency [40].

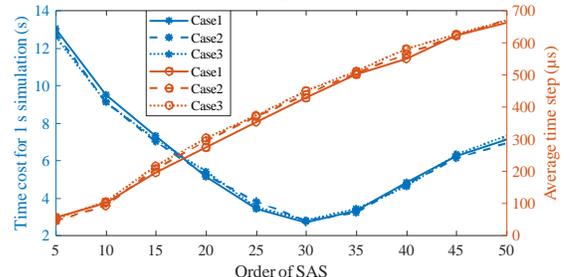

Fig. 10. Performance of SAS with different orders on the 39-bus system

The results presented in Fig. 10 demonstrate that as the SAS order increases, the time step grows, and the total time cost first decays and then rises due to the increased computation per step. The optimal order is around 30, with approximately a time cost of 2.75 s for a 1-s simulation on the 39-bus system.

### C. Comparison of SAS with Traditional Numerical Methods

In Table II below, the performance of the 30[th]-order SAS is then compared with numerical solvers including the Numerical Differentiation Formulas (NDF) method, Runge-Kutta method, and Trapezoidal-rule method, which are provided by the MATLAB solvers *ode*15*s*, *ode*45, and *ode*23*t*, respectively. Also, the auxiliary '*odeset*' function in MATLAB with a carefully adjusted absolute error tolerance around $10^{-5}$ is used to generate results with per unit bus voltage errors less than 0.01 p.u. Because the MATLAB solvers mentioned above all adopt variable time steps, the average time steps are calculated for comparisons. As demonstrated in Table II, the 30[th]-order SAS offers a significant advantage in terms of time efficiency, thanks to its capacity to leverage large time steps enabled by high-order approximations.

TABLE II
Comparing Time Steps and Time Costs of Different Methods

| Case | NDF (ode15s) | | Runge-Kutta (ode45) | | SAS (30[th]-order) | | Trapezoidal rule (ode23t) | |
|---|---|---|---|---|---|---|---|---|
| | Time step (μs) | Time cost (s) | Time step (μs) | Time cost (s) | Time step (μs) | Time cost (s) | Time step (μs) | Time cost (s) |
| Case 1 | 45.8 | 17.2 | 50.5 | 13.5 | 469 | 2.73 | 48.3 | 15.7 |
| Case 2 | 46.2 | 17.4 | 50.3 | 13.6 | 464 | 2.71 | 48.6 | 15.5 |
| Case 3 | 46.1 | 17.3 | 50.7 | 13.3 | 467 | 2.81 | 48.9 | 15.9 |

Fig. 11 shows a comparison of the varying time steps and maximum absolute bus voltage errors for Case 2, from which the 30[th]-order SAS can gradually increase the time step to around five times of the initial step, and its simulation result is more accurate than the results from other solvers.

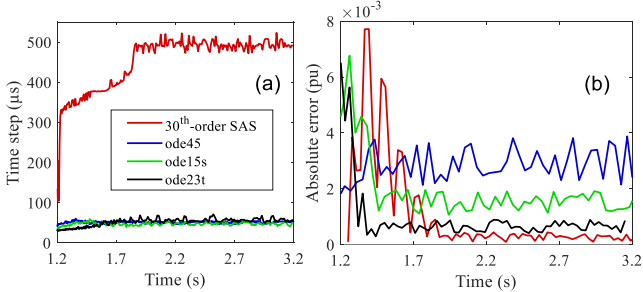

Fig. 11. Comparison of time step and maximum absolute error between the SAS approach and traditional numerical approaches

### D. Comparison of SAS with PSCAD

Conducting PSCAD simulations for the three cases. The average time costs for 1-s simulations, as well as the maximum and average errors of three-phase voltages at different time steps are summarized in Table III, where the simulation results using a 1 μs time step is designated as the benchmark. Remarkably, all three cases have nearly identical time cost using the same time step. The maximum error is the maximum of three-phase voltage absolute error of the 39 buses, and average error is the average of all voltage errors related to the benchmark during $t$=0 to 2 s.

Fig. 12 and Fig. 13 respectively illustrate the simulated phase-C voltage at Bus 10 and its error for Case 2 using different time steps. From Fig. 13, the PSCAD results converge to the benchmark results shortly, but obvious errors exist for fast dynamics following a disturbance with times steps >50 μs. The reason is that the trapezoidal-rule method based nodal approach employed in PSCAD [41] has a low order and less accuracy without many iterations.

TABLE III
Comparison of Performance on the IEEE 39-Bus System

| Approach | PSCAD | | | | SAS |
|---|---|---|---|---|---|
| Time step (μs) | 5 | 50 | 75 | 100 | 467 |
| Time cost (s) | 56.2 | 5.62 | 3.75 | 2.81 | 2.71 |
| Maximum error (pu) | 1.10 | 5.52 | 6.11 | 5.95 | 7.7×10⁻³ |
| Average error (×10⁻³ pu) | 0.15 | 1.73 | 2.80 | 4.11 | 0.032 |

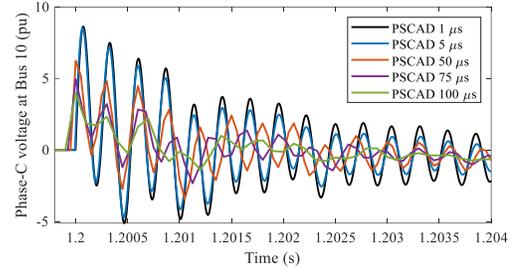

Fig. 12. Phase-C voltages of Bus 10 from PSCAD

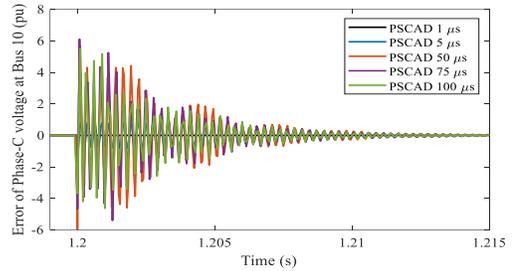

Fig. 13. Errors of phase-C voltages at Bus 10 from PSCAD

In comparison, the results provided by the SAS-based approach are presented in Fig. 14. Utilizing a high order approximation, the SAS approach could provide accurate results for fast dynamics under large time steps, and detailed high resolution dynamics can be reconstructed through dense output.

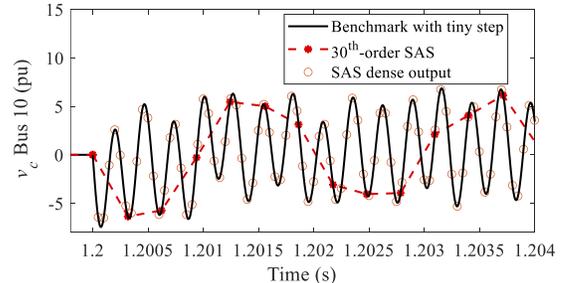

Fig. 14. Phase-C voltage at Bus 10 simulated by SAS

To compare simulation using the SAS approach with a 467-µs time step, PSCAD performs faster when its time step increases to over 100 µs. However, PSCAD has much bigger errors than the SAS approach. Even if PSCAD uses a 5-µs time step, its average and maximum errors are respectively 4.7 and 142.9 times of those with the SAS approach using the 467-µs time step. In such a case, the time cost of PSCAD is 20.7 times that with the SAS approach. Thus, when both high accuracy and speed of simulation are desired, the SAS approach is superior.

### E. Test of Limit Violation Detection Algorithm

This test activates the limits on $E_{fd}$ of exciters and $p_1$ of governors and re-conducts Case 2 to validate the proposed switch detection algorithm.

In this scenario, the $E_{fd}$ of SG4 that connected to Bus 34 reaches its upper limit. The simulation results provided by the $30^{th}$-order SAS with and without the proposed switch detection algorithm are compared against the benchmark results, as shown in Fig. 15. Without detection of the state switch moments, variable values that exceed their limits are not forced to be the limits immediately. The resulting trajectory during the interval of that time step is inaccurate and may compromise the accuracy of the trajectory in subsequent intervals. In contrast, the proposed detection algorithm accurately locates the switching moment and corrects the result immediately within <1 µs, ensuring an accurate result matching the benchmark.

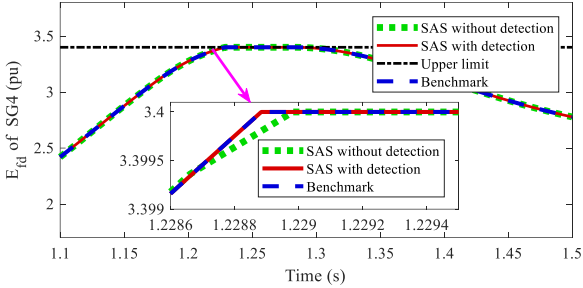

Fig. 15. Simulation results of $E_{fd}$ of SG4

## VII. Case Study on a 390-Bus System

To test the time performance of the SAS-based EMT simulation approach on a large system, a synthetic 390-bus system is constructed by interconnecting 10 replications of the 39-bus system [42], in which generators are replaced by ideal current sources.

Subsequently, by simulating a grounding fault at $t$=1 s lasting for 0.2 s, the performance of the SAS-based simulation is compared with that of the nodal formulation approach [41] utilized in PSCAD. The results are presented in Table IV below, where the time costs are for a 1-second simulation.

TABLE IV
Comparison of Performance on a 390-Bus System

| Approach | PSCAD | | | | SAS |
|---|---|---|---|---|---|
| Time step (µs) | 5 | 50 | 75 | 100 | 534 |
| Time cost (s) | 425 | 42.5 | 28.3 | 21.5 | 98 |
| Maximum error (pu) | 0.22 | 2.58 | 3.36 | 3.85 | $7\times10^{-4}$ |
| Average error ($\times10^{-3}$ pu) | 0.41 | 2.61 | 2.91 | 3.3 | 0.0012 |

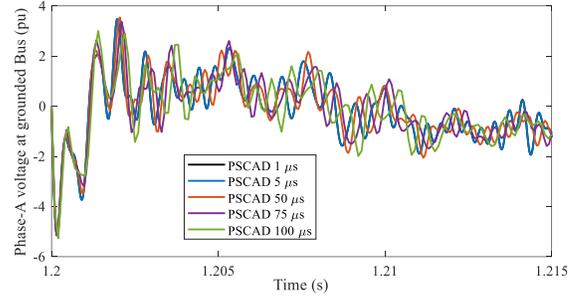

Fig. 16. Phase-A voltage at grounded Bus simulated by PSCAD

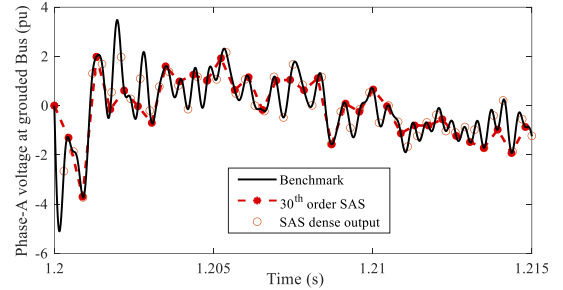

Fig. 17. Phase-A voltage at grounded Bus simulated by the SAS approach

Simulation results of the phase-A voltage at the grounding Bus are shown in Fig. 16-17. From results presented in Table IV and Fig. 16-17, the modal approach in PSCAD has higher efficiency than the state-space SAS approach when adopting a time step approximately >22 µs, which is much smaller than the corresponding time step approximately 100 µs on the 39-bus system. This is aligned with the existing conclusion that the nodal approach has outstanding efficiency on large systems [3].

Nevertheless, same as what is observed on the 39-bus system, the nodal approach failed to accurately capture some fast dynamics right after a disturbance, under a time step larger than 5 µs. In contrast, the SAS-based approach could still provide high precision results with better efficiency.

## VIII. Conclusions

This paper proposed an SAS-based EMT simulation approach to enlarge the time step and reduce computational cost using conventional state-space representation. The incorporated variable time step strategy in the multistage simulation strategy can enhance its robustness, efficiency, and accuracy. Also, the proposed dense output mechanism can provide detailed dynamics under large time steps, and the proposed switch detection algorithm can accurately locate the time instants when limits are violated, making it flexible and adaptable to accurately simulate state switches caused by limit violations. Case studies on the IEEE 39 bus system and a synthetic 390-ubs system revealed that, when sacrificing the accuracy of detailed dynamics such as overvoltage which happen right after a disturbance, the nodal formulation can be more efficient than the proposed SAS-based approach. Conversely, the SAS-based approach has better efficiency for providing high precision results.

In the meanwhile, when the proposed SAS approach is applied to a system or component with stronger stiffness such as detailed power electronics with a high switching frequency, the simulation time step of the proposed SAS-based approach may be constrained to a small value, which could undermine its potential benefits. Such issues will be explored in our future work.